%% file: Template.tex
\documentclass[a4paper]{article}
\usepackage{spconf}

    \usepackage{cite}
    
    \usepackage{scrextend}
    \usepackage{footnote}
    \makesavenoteenv{minpage}
    \makesavenoteenv{itemize}
    \makesavenoteenv{enumerate}
    
    \usepackage{hyperref}
    \newcommand{\figref}[1]{Figure~\ref{#1}}
    \newcommand{\subfigref}[1]{(\protect\subref{#1})}
    \newcommand{\secref}[1]{Section~\ref{#1}}
    \newcommand{\itemref}[1]{\ref{#1}}
    
    \usepackage{amsfonts}
    \usepackage{mathtools}
    \usepackage{amssymb}
    \usepackage{stmaryrd}
    \usepackage{gensymb}
    
    \usepackage{graphicx}
    \usepackage{color}
    \usepackage{subcaption}
    
    \usepackage{pgfplots}
    \usepackage{tikz}
    \usepackage{tkz-graph}
    \usetikzlibrary{shapes, arrows, graphs}
    \newcommand{\colorbar}[3]{\begin{tikzpicture}\begin{axis}[hide axis, scale only axis, height=0pt, width=0pt, colorbar, point meta min=#1, point meta max=#2, colorbar style={height=#3, ytick={#1, ..., #2}}]\end{axis}\end{tikzpicture}}
    
    \usepackage{enumitem}
    
    \let\oldMinipage\minipage \def\minipage{\noindent\oldMinipage}
    \let\oldFigure\figure \def\figure{\oldFigure[!h]\centering}
    \let\oldAlgorithm\algorithm \def\algorithm{\oldAlgorithm[!h]}
    \let\oldEnumerate\enumerate \def\enumerate{\oldEnumerate[label=\alph*)]}
    \newenvironment{subFigure}[1]{\hfill\begin{subfigure}{#1}\centering}{\end{subfigure}\hfill}
    \let\oldInput\input \renewcommand{\input}[1]{\resizebox{\hsize}{!}{\oldInput{#1}}}
    \newenvironment{remark}{\emph{Remark:}}{}

    \newcommand{\set}[1]{\text{$\mathcal{#1}$}}
    \newcommand{\graph}[1]{\text{$\mathcal{#1}$}}
    \renewcommand{\vector}[1]{\textbf{#1}}
    
    \newcommand{\spectral}[1]{\text{$\widehat{#1}$}}
    \renewcommand{\matrix}[1]{\textbf{#1}}
    \newcommand{\matrixCell}[3]{\text{$\matrix{#1}_{\node{#2}, \node{#3}}$}}
    \newcommand{\node}[1]{\text{$#1$}}
    \newcommand{\edge}[2]{\text{$\left(\node{#1}, \node{#2}\right)$}}

    \newcommand{\squaredTimeSpread}{\text{$\Delta_t^2$}}
    \newcommand{\squaredFrequencySpread}{\text{$\Delta_\omega^2$}}

    \newcommand{\squaredSpectralSpread}{\text{$\Delta_s^2$}}
    \newcommand{\squaredGraphSpread}[2]{\text{$\Delta_{\graph{#1}, \node{#2}}^2$}}
    \newcommand{\squaredDistance}[1]{\text{$d_{#1}^2$}}
    \newcommand{\distance}[1]{\text{$d_{#1}$}}
    
    \newcommand{\normalizedLaplacian}[1]{\text{$\matrix{\L}_{\matrix{#1}}$}}
    \newcommand{\laplacian}[1]{\text{$\matrix{L}_{\matrix{#1}}$}}
    \newcommand{\inverse}[1]{\text{$\bar{\text{#1}}$}}
    \newcommand{\eigenValues}[1]{\text{$\Lambda_{#1}$}}
    \newcommand{\eigenVectors}[1]{\text{$\mathcal{X}_{#1}$}}
    \newcommand{\eigenValue}{\text{$\lambda$}}
    \newcommand{\eigenVector}{\text{$\boldsymbol{f}$}}
    \newcommand{\uncertaintyCurve}[1]{\text{$\gamma_{\node{#1}}$}}
    \newcommand{\dirichletForm}[3]{\text{$S_{\graph{#1}, #2}(\vector{#3})$}}
    
    \DeclareMathOperator*{\mathDiag}{\mathrm{diag}}

    \newcommand{\ie}{\emph{i.e }}
    \newcommand{\st}{\text{ s.t. }}
    
    \newcommand{\R}{\text{$\mathbb{R}$}}

\title{Toward An Uncertainty Principle For Weighted Graphs}
\twoauthors
  {Bastien Pasdeloup, R\'eda Alami, Vincent Gripon\sthanks{This work was supported by the European Research Council under the European Union's Seventh Framework Programme (FP7/2007-2013) / ERC grant agreement n\degree~290901.}}
	{Telecom Bretagne\\
	UMR CNRS Lab-STICC\\
	name.surname@telecom-bretagne.eu}
  {Michael Rabbat}
	{McGill University\\
	ECE dept.\\
	name.surname@mcgill.ca}
\begin{document}

\maketitle
            \begin{abstract}
                
                The uncertainty principle states that a signal cannot be localized both in time and frequency.
                With the aim of extending this result to signals on graphs, Agaskar \& Lu \cite{Agaskar2012} introduce notions of graph and spectral spreads.
                They show that a graph uncertainty principle holds for some families of unweighted graphs.
                This principle states that a signal cannot be simultaneously localized both in graph and spectral domains.
                In this paper, we aim to extend their work to weighted graphs.
                We show that a naive extension of their definitions leads to inconsistent results such as discontinuity of the graph spread when regarded as a function of the graph structure.
                To circumvent this problem, we propose another definition of graph spread that relies on an inverse similarity matrix.
                We also discuss the choice of the distance function that appears in this definition.
                Finally, we compute and plot uncertainty curves for families of weighted graphs.

            \end{abstract}

            
            \section{Introduction}
            \label{intro}
                
                In classical signal processing holds an uncertainty principle stating that a signal cannot be localized both in time and frequency domains \cite{Folland1997}.
                This tradeoff is defined by the following equation
                    \begin{equation}
                        \squaredTimeSpread \squaredFrequencySpread \geq \frac{1}{4}
                        \label{classicalUncertainty}
                    \end{equation}
                    in which \squaredTimeSpread{} is the \emph{time spread} of the signal and \squaredFrequencySpread{} its \emph{frequency spread}.
                
                Graph signal processing \cite{Shuman2013a} is a generalization of classical Fourier analysis in which the support for the signal is not necessarily a uniform sampling in time but may be a more complex structure, represented as a graph.
                This emerging domain has received a lot of interest recently \cite{Hammond2011, Narang2011, Shuman2013b} and has been applied to fields such as image denoising \cite{Shuman2013a} and social networks \cite{Rabbat2014}.
                
                In the context of signal processing on graphs, \cite{Agaskar2012} introduces a spectral graph uncertainty principle analog to \eqref{classicalUncertainty}, stating that a signal on a graph cannot be localized both in the graph domain and in the spectral domain.
                For a given signal, the authors propose notions of \emph{graph spread} around a node \node{u_0}, that we denote by \squaredGraphSpread{G}{u_0}, and \emph{spectral spread} around frequency 0, that we denote by \squaredSpectralSpread.
                Note that the choice to consider spectral spread around 0 makes sense for diffusion of signals on graphs, which in most cases converge to a signal aligned with first eigenvalue of the Laplacian.
                They show that for a fixed node \node{u_0} and any signal \vector{x} on a graph, $(\squaredGraphSpread{G}{u_0}(\vector{x}), \squaredSpectralSpread(\vector{x}))$ is higher than a certain curve called \emph{uncertainty curve}.
                The authors then plot this curve for some particular unweighted graphs for which an equation can be determined, and propose an efficient algorithm to estimate it for any unweighted graph.

                In this paper, we aim to extend the results of~\cite{Agaskar2012} to weighted graphs.
                We first review the uncertainty principle for unweighted graphs in \secref{uncertaintyUnweighted}.
                Then, we show in \secref{uncertaintyWeighted} that a naive use of the method introduced in \cite{Agaskar2012} leads to inconsistent results when applied to weighted graphs, and propose a new definition for the graph spread.
                Additionally, we discuss in \secref{examplesOfDistances} the choice of the distance function that appears in our definition of graph spread.
                Finally, in \secref{results}, we use our definition to plot uncertainty curves for some weighted graphs, using various distance functions, and conclude in \secref{conclusion}.
                
            %
        
        
            \section{Uncertainty principle for unweighted graphs}
            \label{uncertaintyUnweighted}
                
                \subsection{Context and definitions}
                \label{context}    
                
                    In this document, we consider a connected, simple graph $\graph{G} = (\set{V}, \set{E}, \matrix{W})$ composed of a set of $|\set{V}| = N$ nodes, a set of edges \set{E}, and a matrix \matrix{W}.
                    Without loss of generality, we label the nodes using integers (\ie $\set{V} = \left\{1 \dots N \right\}$).
                    In the definition of \graph{G}, \matrix{W} is a symmetric matrix of real values such that \matrixCell{W}{u}{v} denotes the weight associated with edge $\edge{u}{v} \in \set{E}$.
                    In the particular case of unweighted graphs, \matrix{W} is the binary adjacency matrix of \graph{G}.
                    
\begin{figure}
                        \begin{subFigure}{0.7\linewidth}
                            \input{graphWithSignal.tex}
                        \end{subFigure}
                        \begin{subFigure}{0.2\linewidth}
                            \colorbar{-1}{1}{2.5cm}
                        \end{subFigure}
                        \caption
                        {
                            Example of graph carrying a signal \vector{x}.
                            The value of \vector{x} associated with each node is described by a color according to the given scale.
                        }
                        \label{graphWithSignal}
                    \end{figure}
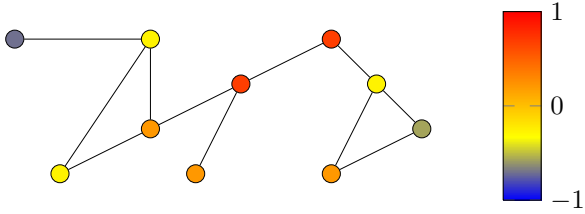

                    A signal \vector{x} on a graph \graph{G} is a set of real values associated with the nodes of \set{V}.
                    Mathematically, $\vector{x} = \left\{\vector{x}(1) \dots \vector{x}(N) \right\}$ is a vector of $\R^N$.
                    \figref{graphWithSignal} depicts an example of graph carrying a signal.
                    
                    A signal \vector{x} is said to be smooth on a graph \graph{G} if nearby nodes carry similar values of signal.
                    Such a measure of smoothness is given by the discrete $p$-Dirichlet form \cite{Shuman2013a} of the signal:
                        \begin{equation}
                            \resizebox{0.85\hsize}{!}
                            {$
                                \dirichletForm{G}{p}{x} \triangleq \frac{1}{p} \sum\limits_{\node{u} \in \set{V}} \left( \sum\limits_{\node{v} \in \set{V} \st \edge{u}{v} \in \set{E}} \matrixCell{W}{u}{v} (\vector{x}(\node{v}) - \vector{x}(\node{u}))^p \right)^{\frac{1}{p}}
                            $}
                            \;.
                            \label{dirichletForm}
                        \end{equation}
                    A smooth signal is associated with a low \dirichletForm{G}{p}{x} value.
                    In particular, $\dirichletForm{G}{p}{x} = 0$ if and only if \vector{x} is constant.

                    When interpreting the significance of \graph{G} with respect to a signal \vector{x}, \eqref{dirichletForm} gives us that \emph{\matrix{W} is analogous to a similarity between nodes}, with the noticeable exception of $\matrixCell{W}{u}{u} = 0$.
                    More generally, a zero value in \matrix{W} indicates the absence of an edge in \graph{G}.
                    As a consequence \set{E} is redundant with \matrix{W} and can be dropped from the definition of \graph{G}.

                    The normalized Laplacian \normalizedLaplacian{W} of \matrix{W} \cite{Chung1997} is a difference operator analogous to the Laplacian operator arising for example in the study of heat diffusion, wave propagation, and harmonic analysis.
                    It is defined by
                        \begin{equation}
                            \normalizedLaplacian{W} \triangleq \matrix{I} - \matrix{D}^{-\frac{1}{2}} \matrix{W} \matrix{D}^{-\frac{1}{2}}
                            \label{normalizedLaplacian}
                        \end{equation}
                        where \matrix{D} is the diagonal matrix of nodes degrees.
                    Since \matrix{D} and \matrix{W} are both symmetric real matrices, \normalizedLaplacian{W} can be diagonalized and described by its orthonormal eigenvectors $\eigenVectors{\normalizedLaplacian{W}} = (\eigenVector_1 \dots \eigenVector_N)$ and associated eigenvalues $\eigenValues{\normalizedLaplacian{W}} = (\eigenValue_1 \leq \dots \leq \eigenValue_N)$.

                \subsection{Notions of spreads for unweighted graphs}
                \label{spreadsUnweighted}    
                    
                    The notions of graph and spectral spreads introduced in this paper are an extension of~\cite{Agaskar2012}.
                    In the following paragraphs we recall their definitions.
                    
                    The graph spread $\squaredGraphSpread{G}{u_0}(\vector{x})$ of a signal \vector{x} around a given node \node{u_0} is defined by
                        \begin{equation}
                            \squaredGraphSpread{G}{u_0}(\vector{x}) \triangleq \frac{1}{\|\vector{x}\|_2^2} \sum\limits_{\node{u} \in \set{V}} \squaredDistance{geo(\matrix{W})}(\node{u_0}, \node{u}) \vector{x}(\node{u})^2
                            \label{graphSpread}
                        \end{equation}
                        where $\vector{x}(\node{u})$ is the value of \vector{x} at node \node{u}, and $\squaredDistance{geo(\matrix{W})}(\node{u_0}, \node{u})$ is the squared geodesic distance -- shortest path -- between \node{u_0} and \node{u} using weights matrix \matrix{W}.
                    Informally, this definition of $\squaredGraphSpread{G}{u_0}(\vector{x})$ quantifies the distance from \node{u_0} to signal \vector{x}.
                    It allows us to introduce a notion of locality of the signal in \graph{G}: the smaller the graph spread is, the more \vector{x} is concentrated around \node{u_0}.
                   
                    The spectral spread $\squaredSpectralSpread(\vector{x})$ of \vector{x} is defined by
                        \begin{equation}
                            \squaredSpectralSpread(\vector{x}) \triangleq \frac{1}{\|\vector{x}\|_2^2} \sum\limits_{n = 1}^{N} \eigenValue_n \spectral{x}_n^2
                            \label{spectralSpread}
                        \end{equation}
                        where $\spectral{\vector{x}} = (\spectral{x}_1 \dots \spectral{x}_N) \triangleq (\eigenVector_{1}^\top \vector{x} \dots \eigenVector_{N}^\top \vector{x})$ is the graph Fourier transform \cite{Shuman2013a} of \vector{x}.
                    
\begin{figure}
                        \Large
                        {
                            \input{exampleUncertaintyCurves.tex}
                            \caption
                            {
                                Examples of uncertainty curves for some unweighted graphs of $100$ nodes.
                                For the star graph, the middle node (\ie the node connected to all others) is chosen as \node{u_0}.
                            }
                            \label{exampleUncertaintyCurves}
                        }
                    \end{figure}
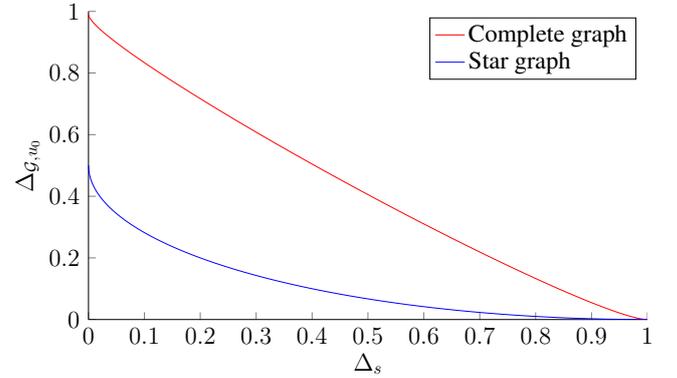

                    One can show that for any signal \vector{x} on an unweighted graph \graph{G}, there exists a relation between \eqref{graphSpread} and \eqref{spectralSpread} such that any pair $(\squaredGraphSpread{G}{u_0}(\vector{x}), \squaredSpectralSpread(\vector{x}))$ is constrained from below by a certain curve \uncertaintyCurve{u_0}.
                    \figref{exampleUncertaintyCurves} depicts the uncertainty curve for some chosen graphs of $100$ nodes.
                    Additional examples of uncertainty curves are proposed in \cite{Agaskar2012}.
                    
                    It is shown in \cite{Agaskar2012} that any uncertainty curve intersects the horizontal axis at exactly one location $(1, 0)$ obtained for a signal \vector{x} localized at node \node{u_0}.
                    Moreover, the curve reaches a spectral spread of $0$ for $\vector{x} = \eigenVector_1$ \cite{Agaskar2012}.
                    In this case, the associated graph spread is equal to $\eigenVector_1^\top \matrix{P}^2 \eigenVector_1$, where $\matrix{P} = \mathDiag\limits_{\node{u} \in \set{V}}(\distance{geo(\matrix{W})}(\node{u_0}, \node{u}))$.
                    
                    In the remainder of this document, we consider unit-norm signals to simplify the reasoning.
                    Therefore, \eqref{graphSpread} becomes
                        \begin{equation}
                            \squaredGraphSpread{G}{u_0}(\vector{x}) \triangleq \sum\limits_{\node{u} \in \set{V}} \squaredDistance{geo(\matrix{W})}(\node{u_0}, \node{u}) \vector{x}(\node{u})^2
                            \label{graphSpreadUnitNorm}
                        \end{equation}
                        and \eqref{spectralSpread} becomes
                        \begin{equation}
                            \squaredSpectralSpread(\vector{x}) \triangleq \sum\limits_{n = 1}^{N} \eigenValue_n \spectral{x}_n^2
                            \;.
                            \label{spectralSpreadUnitNorm}
                        \end{equation}
                
                %
                
            %
        
        
            \section{Towards an uncertainty principle for weighted graphs}
            \label{uncertaintyWeighted}
                
                In this section we aim to extend the definitions of~\cite{Agaskar2012} to weighted graphs.

                In the next subsection, we show that a naive use of \eqref{graphSpreadUnitNorm} leads to inconsistent results such as discontinuity of the graph spread when regarded as a function of \graph{G}.
            
                \subsection{Discontinuity of the graph spread for weighted graphs}
                \label{discontinuityGraphSpread}
                    
                    Let us consider the graph \graph{G} in \figref{problemGeodesic} in which \node{u_0} is fixed and \vector{x} is equally distributed among the nodes.
                    
                    \begin{figure}
                        \begin{subFigure}{0.5\linewidth}
                            \resizebox{0.99\hsize}{!}
                            {
                                \input{problemGeodesic.tex}
                            }
                            \caption{}
                            \label{problemGeodesicGraph}
                        \end{subFigure}
                        \begin{subFigure}{0.4\linewidth}
                            \resizebox{0.7\hsize}{!}
                            {
                                \bordermatrix
                                {
                                    ~          & \node{u_0}  & \node{u}    & \node{v} \cr
                                    \node{u_0} & 0           & \varepsilon & 1        \cr
                                    \node{u}   & \varepsilon & 0           & 2        \cr
                                    \node{v}   & 1           & 2           & 0        \cr
                                }
                            }
                            \caption{}
                            \label{problemGeodesicMatrix}
                        \end{subFigure}
                        \caption
                        {
                            Example weighted graph \subfigref{problemGeodesicGraph} for which we want to compute the uncertainty curve, and associated matrix of weights \subfigref{problemGeodesicMatrix}.
                            We consider a signal \vector{x} equally reparted on the nodes (\ie $\forall \node{w} \in \set{V} : \vector{x}(\node{w}) = \frac{1}{\sqrt{3}}$).
                        }
                        \label{problemGeodesic}
                    \end{figure}
                    
                    Using \eqref{graphSpreadUnitNorm}, we obtain that $\squaredGraphSpread{G}{u_0}(\vector{x}) = \squaredDistance{geo(\matrix{W})}(\node{u_0}, \node{u}) \vector{x}(\node{u})^2 + \squaredDistance{geo(\matrix{W})}(\node{u_0}, \node{v}) \vector{x}(\node{v})^2 = \frac{\varepsilon^2 + 1}{3} \underset{\varepsilon \to 0}{\longrightarrow} \frac{1}{3}$.
                    It seems reasonable to expect that as $\varepsilon$ tends to $0$, \squaredGraphSpread{G}{u_0} tends to the limit case where $\varepsilon = 0$.
                    In particular, \squaredGraphSpread{G}{u_0} should be robust to measurement noise in scenarios where \matrix{W} is not perfectly known.
                    
                    \figref{limitProblemGeodesic} depicts the matrix of weights associated with the limit graph \graph{G'}.
                    
                    \begin{figure}
                        \begin{subFigure}{0.5\linewidth}
                            \resizebox{0.99\hsize}{!}
                            {
                                \input{limitProblemGeodesic.tex}
                            }
                            \caption{}
                            \label{limitProblemGeodesicGraph}
                        \end{subFigure}
                        \begin{subFigure}{0.4\linewidth}
                            \resizebox{0.7\hsize}{!}
                            {
                                \bordermatrix
                                {
                                    ~          & \node{u_0} & \node{u} & \node{v} \cr
                                    \node{u_0} & 0          & 0        & 1        \cr
                                    \node{u}   & 0          & 0        & 2        \cr
                                    \node{v}   & 1          & 2        & 0        \cr
                                }
                            }
                            \caption{}
                            \label{limitProblemGeodesicMatrix}
                        \end{subFigure}
                        \caption
                        {
                            Matrix of weights \subfigref{limitProblemGeodesicMatrix} representing the limit of \figref{problemGeodesicMatrix} when $\varepsilon \longrightarrow 0$, and associated graph \graph{G'} \subfigref{limitProblemGeodesicGraph}.
                            The edge \edge{u_0}{u} has been removed since $\matrixCell{W}{u_0}{u} = 0$.
                        }
                        \label{limitProblemGeodesic}
                    \end{figure}
                    
                    Again, we use \eqref{graphSpreadUnitNorm} to compute the graph spread for \graph{G'} around \node{u_0}.
                    With this graph, we obtain that $\squaredGraphSpread{G'}{u_0}(\vector{x}) = \frac{10}{3}$, leading to a discontinuity of $\graph{G} \mapsto \squaredGraphSpread{G}{u_0}$.
                    
                    \begin{remark}
                        Looking closely at the above mentioned example, we point out that there is a misuse of \matrix{W} in the definition of \squaredGraphSpread{G}{u_0}.
                        As a matter of fact \eqref{dirichletForm} gives us that \matrix{W} is a similarity matrix, whereas \eqref{graphSpreadUnitNorm} uses it as a distance matrix.
                        More generally we expect the graph spread to grow with the distance between nodes in a graph, that is to say as the similarity decreases.
                        In the next subsection we propose a generic framework for a rectified definition of the graph spread in the case of weighted graphs.
                    \end{remark}

                \subsection{Expected behavior of a graph spread}
                \label{expectedBehaviorGraphSpread}
                    
                    In order to define a new notion of graph spread that does not lead to unexpected behaviors as in \secref{discontinuityGraphSpread}, we present some desired properties on \squaredGraphSpread{G}{u_0}.
                    
                    We expect from a graph spread notion that it captures the locality of a signal \vector{x} in the graph domain.
                    In other words, for a fixed node \node{u_0}, the graph spread around \node{u_0} should measure the extent to which the signal \vector{x} is concentrated around \node{u_0}.
                    To achieve this, we would like to ensure the following properties:
                    \begin{itemize}
                        \item $\squaredGraphSpread{G}{u_0}(\vector{x})$ should be small if \vector{x} is localized around \node{u_0}, and should increase as the distance between \node{u_0} and the nodes carrying \vector{x} increases.
                        \item Additionally, the only situation leading to $\squaredGraphSpread{G}{u_0}(\vector{x}) = 0$ should be when the signal is entirely localized on $u_0$ or nodes that are \emph{indistinguishable} from $u_0$.
                        \item A third desired property is that the graph spread should be \emph{similar} for graphs with \emph{similar} weights (continuity of $\graph{G} \mapsto \squaredGraphSpread{G}{u_0}$).
                    \end{itemize}
                    
                    Moreover it appears to us that the choice of the geodesic distance in \eqref{graphSpreadUnitNorm} is arbitrary.
                    In order to be compliant with the previously enumerated properties, we characterize the class of acceptable functions \distance{}:
                    \begin{enumerate}
                        \item \label{prop1} $\forall \node{u}, \node{v} \in \set{V} : \distance{}(\node{u}, \node{v}) \geq 0$.
                        \item \label{prop2} $\forall \node{u}, \node{v} \in \set{V} : \distance{}(\node{u}, \node{v}) = 0 \Leftrightarrow \forall \node{w} \in\set{V} : \distance{}(\node{u}, \node{w}) = \distance{}(\node{v}, \node{w})$.
                        \item \label{prop3} \distance{} is continuous, and if we increase \matrixCell{W}{u}{v} for a single edge \edge{u}{v}, then $\forall \node{u'}, \node{v'} \in \set{V} : \distance{}(\node{u'}, \node{v'})$ does not increase.
                    \end{enumerate}

                    \begin{remark}
                        The geodesic distance \squaredDistance{geo(\matrix{W})} based on \matrix{W} is not compliant with \itemref{prop3} (not continuous and increasing with \matrix{W}).
                    \end{remark}
                    
                %
                
            %
            
        
            \section{Examples of compliant distances for graph spread}                   
            \label{examplesOfDistances}

                In this section we propose two choices of distances compliant with the previously introduced properties.

                \subsection{Inverse similarity matrix}
                \label{inverseSimilarityMatrix}

                    The distance described in this subsection is a simple rectified version of \eqref{graphSpreadUnitNorm} and is compatible with it in the case of unweighted graphs.

                    Let us consider a graph \graph{G}.
                    We introduce a new matrix \matrix{\inverse{S}} as follows:
                        \begin{equation}
                            \forall \node{u}, \node{v} \in \set{V} : \matrixCell{\inverse{S}}{u}{v} \triangleq \left\{
                                                                                                                   \begin{array}{ll}
                                                                                                                       \infty                         & \textbf{if } \matrixCell{W}{u}{v} = 0 \\
                                                                                                                       0                              & \textbf{if } \matrixCell{W}{u}{v} = \infty \\
                                                                                                                       \frac{1}{\matrixCell{W}{u}{v}} & \textbf{otherwise}
                                                                                                                   \end{array}
                                                                                                               \right.
                            \;.
                            \label{similarityMatrix}
                        \end{equation}

                    We propose to use it instead of \matrix{W} in \eqref{graphSpreadUnitNorm}.

                    \begin{remark}
                        The choice of taking the inverse is arbitrary and could be replaced by other functions.
                        Standard alternatives are Gaussian kernels, as shown later in \secref{resultsGaussian}.
                        In some cases weighted similarity graphs are constructed from distance graphs and in such cases it appears more natural to use the latter directly instead of estimating it back from \matrix{W}.
                        Some examples of such graphs are given in the next section.
                    \end{remark}
                        
                    We now show that the squared geodesic distance using \matrix{\inverse{S}}, \squaredDistance{geo(\matrix{\inverse{S}})}, is compliant with the three properties enounced in \secref{expectedBehaviorGraphSpread}:
                        \begin{enumerate}
                            \item is trivially true, since $\squaredDistance{geo(\matrix{\inverse{S}})}(\node{u}, \node{v})$ features a square.
                            \item is ensured for any couple of nodes (\node{u}, \node{v}) being $0$-distant (according to \matrix{\inverse{S}}), as for any node \node{w} the shortest path $\node{w} \rightarrow \dots \rightarrow \node{u}$ can be extended to $\node{w} \rightarrow \dots \rightarrow \node{u} \rightarrow \node{v}$ without changing its length (since we add $0$ to it).
                            \item is in most cases trivial.
                                  The only concern is when an edge is removed from \graph{G}.
                                  Such a scenario occurs in the case where the similarity between two nodes \node{u} and \node{v} becomes zero.
                                  By definition of $\matrix{\inverse{S}}$, this corresponds to a distance between \node{u} and \node{v} that diverges to infinity.
                                  It is obvious that eventually the shortest paths of \matrix{\inverse{S}} do not include this edge.
                        \end{enumerate}
                    
                    With this function, the definition of graph spread in \eqref{graphSpreadUnitNorm} now becomes
                        \begin{equation}
                            \squaredGraphSpread{G}{u_0}(\vector{x}) \triangleq \sum\limits_{\node{u} \in \set{V}} \squaredDistance{geo(\matrix{\inverse{S}})}(\node{u_0}, \node{u}) \vector{x}(\node{u})^2
                            \;.
                            \label{graphSpreadCorrectedUnitNorm}
                        \end{equation}

                \subsection{Diffusion distance}
                \label{diffusionDistance}
                    
                    Another distance function we study in this paper is the diffusion distance, as defined in \cite{Segarra2014}.
                    Given a graph adjacency matrix \matrix{W} and its associated (non-normalized) Laplacian matrix \laplacian{W} \cite{Chung1997}, \distance{diff} is defined in matrix form for some constant parameter $\alpha$ as follows:
                        \begin{equation}
                            \forall \node{u}, \node{v} \in \set{V} : \distance{diff}(\node{u}, \node{v}) \triangleq \| (\matrix{I} + \alpha \laplacian{W})^{-1} (\vector{x}_\node{u} - \vector{x}_\node{v}) \|
                            \label{diffusionDistance}
                        \end{equation}
                        where $\vector{x}_\node{u}$ is a unit-norm signal having only one non-zero value on node \node{u}.
                    
                    One can show that \distance{diff} verifies the three desired properties presented in \secref{expectedBehaviorGraphSpread}.
                    In the remaining of the document, we set $\alpha = 1$ and use the $l_2$ norm.
                    
                %
                
            %

        
            \section{Results for classical weighted graphs}
            \label{results}

                In this section we introduce several classical weighted graphs and plot their uncertainty curves considering both inverse similarity matrix and diffusion distance.
                Curves are plotted using the Sandwich algorithm introduced in \cite{Agaskar2012}.
                By comparing the resulting curves to known uncertainty curves \cite{Agaskar2012} obtained for graphs such as the ring or star graphs, one can evaluate the \emph{amount of uncertainty} associated to the graph under study.
                
                \subsection{Random graph}
                \label{resultsRandom}

\begin{figure}
                        \Large
                        {
                            \input{randomGraphs.tex}
                            \caption
                            {
                                Examples of uncertainty curves for some randomly weighted families of graphs of $10$ nodes.
                                The curves are computed for the two distance functions \distance{geo(\matrix{\inverse{S}})} and \distance{diff}.
                                Mean uncertainty curves for $100$ random weights.
                            }
                            \label{randomGraphs}
                        }
                    \end{figure}
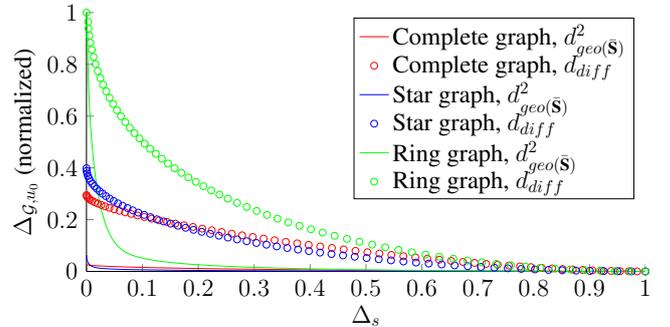

                    We call random graph a graph which adjacency symmetric matrix \matrix{W} is such that each non-null coordinate \matrixCell{W}{u}{v} is drawn uniformly between 0 and 1.
                    Using the previously introduced distance functions, we plot in \figref{randomGraphs} the uncertainty curves for such families of graphs.
                    The curves are normalized such that the graph spread associated with $\squaredSpectralSpread(\vector{x}) = 0$ is at most equal to $1$ for each distance function used.
                    
                    It is interesting to notice that the choice of the distance does not impact the relative order of the curves.
                    Additionally, the intersection between the uncertainty curves associated to the star and complete graphs is kept when switching the distance function.
                    The main difference is the smoothness of the curves.
                    Using \distance{diff} tends to produce uncertainty curves that are more regular than when using \distance{geo(\matrix{\inverse{S}})}.

                \subsection{Gaussian kernel}
                \label{resultsGaussian}

                    We consider graphs obtained using a Gaussian kernel.
                    The idea is to build a distance graph and to apply a Gaussian kernel to all weights to obtain \matrix{W}.
                    The Gaussian kernel has two parameters $\alpha$ and $\beta$ and is defined as follows:
                        \begin{equation}
                            g: x \mapsto \alpha \exp\left(-\beta x^2\right)
                            \;.
                            \label{gaussianKernel}
                        \end{equation}
                    
                    We consider a set \set{S} of $N$ sensors uniformly distributed in a $1 \times 1$ square.
                    We define a symmetric matrix \matrix{E} as follows.
                    Fix some radius $r$ such that if two sensors \node{u} and \node{v} are at Euclidean distance $\distance{euc}(\node{u}, \node{v})$ less than $r$, then $\matrixCell{E}{u}{v} = \frac{\distance{euc}(\node{u}, \node{v})}{\max\limits_{\node{u'}, \node{v'} \in \set{V}} \distance{euc}(\node{u'}, \node{v'})}$ and $\matrixCell{E}{u}{v}=0$ otherwise.
                    \matrix{W} is then defined by applying $g$ to each cell of \matrix{E}.
                    
                    \figref{gaussianKernel} depicts the mean uncertainty curves for random geometric graphs.
                    When computing the uncertainty curve using the squared geodesic distance \squaredDistance{geo(\matrix{E})}, we directly use the matrix of Euclidean distances \matrix{E}, and do not retrieve it from \matrix{W} (see remark in \secref{inverseSimilarityMatrix}).
                    The curves are normalized so that no value of \squaredGraphSpread{G}{u_0} exceeds $1$ for each distance function.
                    
                    \begin{figure}
                        \Large
                        {
                            \input{gaussianKernel.tex}
                            \caption
                            {
                                Uncertainty curves for random geometric graphs of $10$ nodes.
                                Parameters $(\alpha, \beta, r)$ are respectively fixed to $(1, 1, 0.3)$.
                                Mean uncertainty curves for $100$ random graphs.
                            }
                            \label{gaussianKernel}
                        }
                    \end{figure}
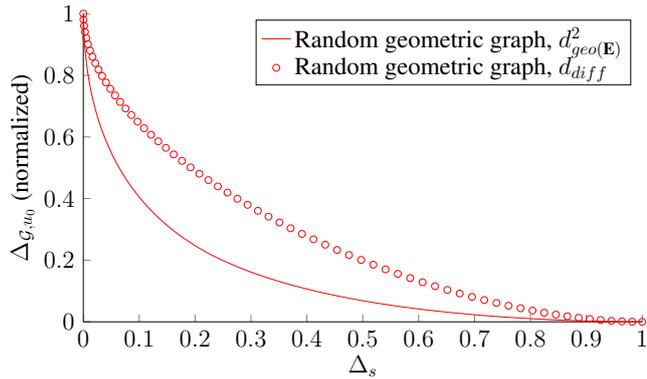
                    
\begin{figure}
                        \begin{subFigure}{0.3\linewidth}
                            \vspace{0.08cm}
                            \includegraphics[width=\textwidth]{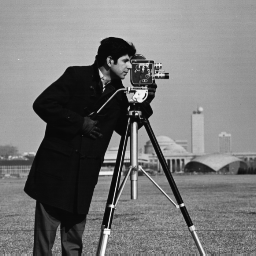}
                            \vspace{0.08cm}
                            \caption{}
                            \label{cameramanImage}
                        \end{subFigure}
                        \begin{subFigure}{0.6\linewidth}
                            \Large
                            {
                                \input{cameraman.tex}
                                \caption{}
                                \label{cameramanCurve}
                            }
                        \end{subFigure}
                        \caption
                        {
                            Computation of the normalized uncertainty curves \subfigref{cameramanCurve} associated to the image \subfigref{cameramanImage}, for the introduced distance functions.
                            Parameters $(\alpha, \beta)$ are respectively fixed to $(1, 1)$.
                        }
                        \label{cameraman}
                    \end{figure}

                    Additionally, we apply the same Gaussian kernel to semi-localized graphs.
                    We use the same graph as presented in \cite{Shuman2013a} (Example 2).
                    Such a graph is obtained by connecting pixels of the $32 \times 32$ cameraman image to their eight neighbors, weighting connections using $g$ over the difference of intensity of pixels to obtain \matrix{W}.
                    This method for constructing graphs for images has been previously used for example in \cite{Narang2012}.
                    \figref{cameramanImage} depicts the picture from which the graph is extracted.
                    \figref{cameramanCurve} shows the associated uncertainty curves using the distances \distance{geo(\matrix{\inverse{S}})}\footnote
                        {
                            Contrary to the study of random geometric graphs, we do not directly use a matrix of distances \matrix{D} associated to the difference of pixels intensity, but retreive \matrix{\inverse{S}} from \matrix{W} using \eqref{similarityMatrix}.
                            As a matter of fact, two adjacent pixels with identical intensity result in a distance of $0$ if using \matrix{D}, and would cause the discontinuity problem previously introduced.
                            A solution to cope with this problem is to add an $\varepsilon$ noise to all edge weights.
                            However, this leads to hard to visualize curves.
                            Therefore, for the sake of comprehension, we use \squaredDistance{geo(\matrix{\inverse{S}})} and not \squaredDistance{geo(\matrix{D})}.
                        }
                        and \distance{diff}.
                    
                %
                
        
            \section{Conclusion}
            \label{conclusion}
                
                In this work, we have extended the notion of uncertainty on graphs introduced in \cite{Agaskar2012} to weighted graphs, and pointed out important properties of the distance function used in the definition of graph spread.
                We have shown the applicability of our work on classical families of graphs, as well as on semi-localized graphs that are encountered in real-life use-cases.
                
                A direction of our future work will focus on side aspects, such as determining a way to efficiently choose the node used as \node{u_0} in the computation of \squaredGraphSpread{G}{u_0} to perform better comparisons of uncertainty curves.
                We will also investigate some properties that could be derived from the uncertainty of a given graph when considering some categories of signals.
                
            %

\bibliographystyle{IEEEbib}
\bibliography{refs}

\end{document}

%% file: graphWithSignal.tex

    \tikzstyle{graphNode} =
    [
        circle,
        draw,
        text centered,
        minimum height=0.4cm
    ]
    
    \tikzstyle{graphEdge} =
    [
        draw
    ]
    
    \definecolor{signalValue1}{RGB}{255, 62, 0}
    \definecolor{signalValue2}{RGB}{255, 152, 0}
    \definecolor{signalValue3}{RGB}{255, 243, 0}
    \definecolor{signalValue4}{RGB}{164, 164, 91}
    \definecolor{signalValue5}{RGB}{112, 112, 143}


    \begin{tikzpicture}
        
        \node[graphNode, fill=signalValue5] at (0, 3) (1)  {};
        \node[graphNode, fill=signalValue3] at (1, 0) (2)  {};
        \node[graphNode, fill=signalValue2] at (3, 1) (3)  {};
        \node[graphNode, fill=signalValue3] at (3, 3) (4)  {};
        \node[graphNode, fill=signalValue2] at (4, 0) (5)  {};
        \node[graphNode, fill=signalValue1] at (5, 2) (6)  {};
        \node[graphNode, fill=signalValue2] at (7, 0) (7)  {};
        \node[graphNode, fill=signalValue1] at (7, 3) (8)  {};
        \node[graphNode, fill=signalValue3] at (8, 2) (9)  {};
        \node[graphNode, fill=signalValue4] at (9, 1) (10) {};
        
        \draw[graphEdge] (1) -- (4);
        \draw[graphEdge] (2) -- (3);
        \draw[graphEdge] (2) -- (4);
        \draw[graphEdge] (3) -- (4);
        \draw[graphEdge] (3) -- (6);
        \draw[graphEdge] (5) -- (6);
        \draw[graphEdge] (6) -- (8);
        \draw[graphEdge] (7) -- (9);
        \draw[graphEdge] (7) -- (10);
        \draw[graphEdge] (8) -- (9);
        \draw[graphEdge] (9) -- (10);
        
    \end{tikzpicture}


%% file: exampleUncertaintyCurves.tex
%
%
\begin{tikzpicture}

\begin{axis}[%
width=4.5in,
height=2.5in,
at={(0.808889in,0.513333in)},
scale only axis,
every outer x axis line/.append style={black},
every x tick label/.append style={font=\color{black}},
xmin=0,
xmax=1,
xlabel={$\Delta_s$},
every outer y axis line/.append style={black},
every y tick label/.append style={font=\color{black}},
ymin=0,
ymax=1,
ytick={0,0.2,0.4,0.6,0.8,1.0},
ylabel={$\Delta_{\mathcal{G}, u_0}$},
axis x line*=bottom,
axis y line*=left,
legend style={legend cell align=left,align=left,draw=black}
]
\addplot [color=red,solid]
  table[row sep=crcr]{%
0	0.989999999999999\\
0.000533254604194261	0.984911566823759\\
0.0021318923493016	0.978799151598007\\
0.00479253741080698	0.971675661838048\\
0.0085095713358487	0.963556140133363\\
0.0132751449076229	0.954457732382414\\
0.0190791947204934	0.944399651585777\\
0.0259094644307734	0.933403137274311\\
0.0337515306382892	0.921491410657888\\
0.0425888333441189	0.908689625589423\\
0.0524027109201077	0.895024815447834\\
0.0631724395164741	0.880525836051881\\
0.0748752768240932	0.865223304725706\\
0.0874865100991064	0.849149535644684\\
0.100979508348628	0.832338471597811\\
0.115325778567035	0.814825612311207\\
0.13049502590426	0.796647939483825\\
0.146455217639234	0.777843838693433\\
0.163172650822991	0.758453018338267\\
0.180612023448489	0.738516425785617\\
0.198736508997752	0.718076160903399\\
0.217507834207973	0.697175387158373\\
0.23688635989259	0.675858240468521\\
0.256831164646906	0.654169736001704\\
0.277300131261592	0.632155673117404\\
0.298250035660475	0.609862538653426\\
0.319636638177118	0.587337408759307\\
0.341414776974625	0.564627849486827\\
0.363538463413423	0.541781816345207\\
0.385960979165314	0.51884755303368\\
0.40863497486779	0.495873489566045\\
0.431512570110807	0.472908140001929\\
0.454545454545457	0.449999999999996\\
0.47768498990118	0.427197444409465\\
0.500882312693047	0.404548625119047\\
0.524088437407679	0.382101369373738\\
0.54725435994471	0.359903078779236\\
0.570331161097846	0.338000629204839\\
0.593270109857974	0.316440271795264\\
0.616022766316805	0.295267535303675\\
0.638541083956988	0.274527129948965\\
0.660777511112617	0.25426285300063\\
0.682685091382277	0.234517496293724\\
0.704217562786244	0.215332755865887\\
0.72532945545795	0.196749143907799\\
0.745976187662286	0.178805903213963\\
0.766114159937644	0.161540924314762\\
0.785700847165034	0.144990665462887\\
0.804694888367506	0.129190075645213\\
0.823056174051646	0.114172520781427\\
0.840745930906831	0.0999697132651789\\
0.857726803681435	0.0866116449981646\\
0.873962934066365	0.0741265240558538\\
0.889420036416141	0.0625407151212329\\
0.904065470149353	0.0518786838108899\\
0.917868308676168	0.0421629450106893\\
0.93079940470494	0.0334140153319049\\
0.942831451792223	0.0256503697864729\\
0.953939042005833	0.0188884027732882\\
0.964098719578199	0.0131423934585414\\
0.973289030437741	0.00842447562248954\\
0.981490567513175	0.00474461203669809\\
0.98868601171516	0.00211057342580772\\
0.99486016850885	0.000527922058151461\\
1	0\\
};
\addlegendentry{Complete graph};

\addplot [color=blue,solid]
  table[row sep=crcr]{%
0	0.5\\
0.000301181303795762	0.487729385738535\\
0.00120454379482829	0.475466162836295\\
0.00270954332130741	0.463217718200188\\
0.00481527332780214	0.450991429835221\\
0.00752046540129404	0.438794662400375\\
0.01082349003522	0.42663476277232\\
0.0147223576110539	0.414519055619867\\
0.019214719596769	0.402454838991936\\
0.0242978699614722	0.390449379921566\\
0.0299687468054551	0.378509910048368\\
0.0362239342045604	0.366643621262551\\
0.0430596642677915	0.354857661372767\\
0.0504718194069693	0.343159129800547\\
0.0584559348169835	0.331555073303883\\
0.0670072011652691	0.320052481732494\\
0.076120467488713	0.308658283817456\\
0.0857902442964595	0.297379342997515\\
0.0960107068765509	0.286222453284865\\
0.10677569880448	0.275194335172703\\
0.11807873565164	0.264301631587006\\
0.129913008891263	0.25355090388513\\
0.142271389999719	0.242948627903395\\
0.155146434750303	0.232501190056445\\
0.168530387697452	0.2222148834902\\
0.182415186848426	0.212095904291068\\
0.196792468519368	0.202150347753774\\
0.211653572373422	0.192384204709669\\
0.226989546637267	0.182803357918176\\
0.242791153493496	0.173413578523123\\
0.259048874645033	0.164220522576495\\
0.275752917048526	0.155229727631471\\
0.292893218813452	0.146446609406726\\
0.310459455262912	0.137876458524277\\
0.328441045152976	0.129524437322522\\
0.346827157046236	0.121395576746752\\
0.365606715836355	0.113494773318631\\
0.384768409419392	0.10582678618669\\
0.404300695507572	0.0983962342596757\\
0.42419180858215	0.0912075934242097\\
0.4444297669804	0.0842651938487266\\
0.46500238011293	0.0775732173751386\\
0.485897255806791	0.0711356949998608\\
0.507101807770217	0.0649565044456434\\
0.528603263174009	0.059039367825821\\
0.550388670345374	0.0533878494022465\\
0.572444906569722	0.0480053534382771\\
0.594758685995039	0.0428951221482284\\
0.617316567634912	0.0380602337443566\\
0.640104963464983	0.0335036005826358\\
0.663110146607771	0.0292279674084909\\
0.686318259601115	0.0252359097034807\\
0.709715322745532	0.0215298321338966\\
0.733287242525074	0.0181119671022836\\
0.757019820096723	0.0149843734027293\\
0.780898759843123	0.0121489349807361\\
0.804909677983869	0.00960735979838488\\
0.829038111239712	0.00736117880552804\\
0.853269525544642	0.00541174501760944\\
0.877589324800774	0.00376023270064548\\
0.901982859670439	0.00240763666390154\\
0.926435436400333	0.00135477166065491\\
0.950932325672581	0.00060227189741379\\
0.975458771477089	0.000150590651897878\\
1	0\\
};
\addlegendentry{Star graph};

\end{axis}
\end{tikzpicture}%

%% file: problemGeodesic.tex

    \tikzstyle{graphNode} =
    [
        circle,
        draw,
        fill=blue!10,
        text centered,
        minimum height=0.7cm
    ]
    
    \tikzstyle{graphEdge} =
    [
        draw
    ]


    \begin{tikzpicture}
        
        \node[graphNode] at (0, 1) (1) {\node{u_0}};
        \node[graphNode] at (4, 1) (2) {\node{u}};
        \node[graphNode] at (2, 0) (3) {\node{v}};
        
        \draw[graphEdge] (1) -- node[below] {$1$}           (3);
        \draw[graphEdge] (1) -- node[above] {$\varepsilon$} (2);
        \draw[graphEdge] (2) -- node[below] {$2$}           (3);
        
    \end{tikzpicture}


%% file: limitProblemGeodesic.tex

    \tikzstyle{graphNode} =
    [
        circle,
        draw,
        fill=blue!10,
        text centered,
        minimum height=0.7cm
    ]
    
    \tikzstyle{graphEdge} =
    [
        draw
    ]


    \begin{tikzpicture}
        
        \node[graphNode] at (0, 1) (1) {\node{u_0}};
        \node[graphNode] at (4, 1) (2) {\node{u}};
        \node[graphNode] at (2, 0) (3) {\node{v}};
        
        \draw[graphEdge] (1) -- node[below] {$1$} (3);
        \draw[graphEdge] (2) -- node[below] {$2$} (3);
        
    \end{tikzpicture}


%% file: randomGraphs.tex
%
%
\begin{tikzpicture}

\begin{axis}[%
width=4.5in,
height=2.1in,
at={(0.911111in,0.513333in)},
scale only axis,
every outer x axis line/.append style={black},
every x tick label/.append style={font=\color{black}},
xmin=0,
xmax=1,
xlabel={$\Delta_s$},
every outer y axis line/.append style={black},
every y tick label/.append style={font=\color{black}},
ymin=0,
ymax=1,
ytick={0,0.2,0.4,0.6,0.8,1.0},
ylabel={$\Delta_{\mathcal{G}, u_0}$ (normalized)},
axis x line*=bottom,
axis y line*=left,
legend style={legend cell align=left,align=left,draw=black}
]
\addplot [color=red,solid]
  table[row sep=crcr]{%
0	0.0270162259271841\\
0.000273700828122728	0.026369171680329\\
0.00109473964255764	0.0257259039346636\\
0.0024629287495254	0.0250865792740593\\
0.00437797370789934	0.0244513490539359\\
0.00683943682640548	0.0238203670032919\\
0.00984685421406355	0.0231937738105599\\
0.0133996481779572	0.0225717134214655\\
0.0174971753919459	0.0219543266459573\\
0.0221382601385284	0.021341776849644\\
0.0273224846062466	0.0207341730868603\\
0.0330490091687394	0.0201316509285846\\
0.0393169544217609	0.0195343435099251\\
0.0461253375480685	0.0189423844453917\\
0.0534732237268733	0.0183559011571693\\
0.0613595855266139	0.0177750220091369\\
0.0697833543982998	0.0171998735219189\\
0.0787458385044228	0.0166304840733772\\
0.0882438252877546	0.0160670481332862\\
0.0982760544857192	0.0155096901070598\\
0.108841142339029	0.0149585354596654\\
0.119937769203428	0.0144137054352639\\
0.131564108475958	0.0138753325641035\\
0.143718338243455	0.0133435462682888\\
0.156398337936634	0.0128184806887376\\
0.169605608960107	0.012300180766525\\
0.183333920819082	0.0117888780311634\\
0.197580249512526	0.0112847185741718\\
0.212340970586716	0.0107878584334083\\
0.22761290239112	0.0102984404011866\\
0.243390510273388	0.00981665489622232\\
0.259668533615783	0.0093426803914297\\
0.276440670765999	0.00887671212700065\\
0.294592622944394	0.00839895434223272\\
0.313277597828074	0.00793023051062786\\
0.332484920682965	0.00747080630344192\\
0.352201723373994	0.00702098182799741\\
0.372416503847883	0.00658102176462503\\
0.393110235644483	0.00615132380469386\\
0.414264950252546	0.0057322474143341\\
0.435859876230085	0.00532418722908124\\
0.45789947162487	0.00492711909117313\\
0.480332927543883	0.00454192969085014\\
0.503133370686585	0.00416906122425622\\
0.526270866056193	0.00380898235733184\\
0.549716949885515	0.00346212217933114\\
0.573434189757655	0.00312901289000198\\
0.597386587079196	0.00281013906717965\\
0.621535281894875	0.00250599666043245\\
0.646048581447041	0.00221469524244571\\
0.670678387890586	0.0019393648279923\\
0.695379461550576	0.00168049320414283\\
0.720104091050711	0.00143856462043467\\
0.744805974369632	0.00121401935456386\\
0.769431947983634	0.00100733405683281\\
0.793930556583074	0.000818931334062455\\
0.818248861878531	0.000649211617490154\\
0.842345837710654	0.00049846005864044\\
0.866155281891178	0.000367113663046282\\
0.889623359733805	0.000255466643921605\\
0.912696387117204	0.00016377263112488\\
0.935321132912853	9.22412492727173e-05\\
0.957446336856583	4.10332499053825e-05\\
0.979021808177752	1.02638114984218e-05\\
1	0\\
};
\addlegendentry{Complete graph, $d_{geo(\bar{\textbf{S}})}^2$};

\addplot [color=red,only marks,mark=o,mark options={solid}]
  table[row sep=crcr]{%
0	0.294990111897342\\
0.000413946859472962	0.290785263611636\\
0.00165556100762821	0.286407854694493\\
0.00372403863035767	0.281860710464935\\
0.00661751915862088	0.277147920260968\\
0.0103338793220723	0.272273220668117\\
0.0148684425915601	0.267242999863534\\
0.020215806864433	0.262063243717497\\
0.0263691015124835	0.256740613497479\\
0.0333253433696907	0.251278220790423\\
0.0410709378603623	0.2456873298894\\
0.0495950493885791	0.239975783736619\\
0.0588854229303227	0.234151822062585\\
0.068928890353463	0.228223746227926\\
0.0797103517164203	0.222200522716574\\
0.0912137893525152	0.216091139580325\\
0.103421917259842	0.20990482046625\\
0.116334993522375	0.203642028521374\\
0.129916614345765	0.197321084699662\\
0.144146029542869	0.190951697122789\\
0.159001395882252	0.184543696150891\\
0.174459955223393	0.178106955773528\\
0.190497799029643	0.17165148824709\\
0.207090180046797	0.165187311959082\\
0.224211461141231	0.158724472829891\\
0.241835270374888	0.15227299684994\\
0.259934281773034	0.145842937363908\\
0.278480523755008	0.139444291599527\\
0.297445358336297	0.13308699941304\\
0.316799459038348	0.126780951937736\\
0.336513071051857	0.120535900369829\\
0.356555858890291	0.114361513276014\\
0.376897066023531	0.108267322820181\\
0.39737555860655	0.102302585298137\\
0.418087330883651	0.09643525379356\\
0.439001233133485	0.0906741973472936\\
0.460085964816946	0.0850280807930804\\
0.481309819147847	0.0795054365823096\\
0.502641530518336	0.0741144333649112\\
0.52404954645291	0.0688630824095159\\
0.545502454209326	0.0637591230780261\\
0.566965495732394	0.0588108598426886\\
0.588411031593155	0.0540246122924815\\
0.609808199978072	0.049407293902369\\
0.631126612055416	0.0449654831602248\\
0.652335600698811	0.0407055878150253\\
0.673406340512566	0.0366333967255237\\
0.694309851953052	0.0327545204376281\\
0.715018031108951	0.0290741772374297\\
0.735421127792916	0.0256116544668148\\
0.755577173090324	0.0223552305737326\\
0.775460763555866	0.0193089979932452\\
0.795047861843262	0.016476619785722\\
0.814313789333079	0.0138616337929482\\
0.833238591012498	0.0114666840136284\\
0.851801775866064	0.00929433070626935\\
0.869984903207848	0.00734669115315905\\
0.887748532822085	0.00562771423962815\\
0.905101132168129	0.00413603084394015\\
0.922029303613798	0.0028727448923025\\
0.938522591692577	0.00183858572879631\\
0.954568427066249	0.00103423274300094\\
0.970165162934636	0.000459638124869438\\
0.9853089813057	0.0001149043951483\\
1	0\\
};
\addlegendentry{Complete graph, $d_{diff}$};

\addplot [color=blue,solid]
  table[row sep=crcr]{%
0	0.0625855970488724\\
7.49996313415215e-05	0.05689967815159\\
0.000300089137914251	0.0515869793231892\\
0.000675561046409116	0.0466450526124984\\
0.00120206172311415	0.0420706728438355\\
0.00188014396058486	0.0378618794218946\\
0.00271174891170252	0.0340123760133976\\
0.00369913852845101	0.0305173749848853\\
0.00484567133706999	0.0273705338876687\\
0.00614733368760539	0.0246224644571409\\
0.00761612804828765	0.0221877366235289\\
0.00925891110871483	0.0200570542166152\\
0.0110842555992508	0.0182163042393556\\
0.0131015359191939	0.0166643729299654\\
0.0153227354268153	0.0153484972862333\\
0.0177610479329473	0.014236213123276\\
0.0204312997323126	0.0132922805669787\\
0.0234318428733283	0.0125164073611294\\
0.0267182767256479	0.0118045091158827\\
0.0303054731373324	0.0111469596444312\\
0.0342067436119104	0.010539189887413\\
0.0384386376365516	0.00997627674363772\\
0.0430129702219471	0.00945602306758651\\
0.0479439444069102	0.00897577227408172\\
0.0532454925575638	0.00853243404113361\\
0.0589479885807269	0.00812205225708307\\
0.06505360805897	0.0077375551699045\\
0.0715780501319155	0.0073759977530994\\
0.078537102024163	0.00703471555386904\\
0.0859490631016146	0.00671128422835513\\
0.0938286600283315	0.00640415672300653\\
0.102192871048114	0.00611172311263364\\
0.111058527339728	0.00583255805085796\\
0.122653535575521	0.00546691866462638\\
0.135152381857773	0.00512190197277746\\
0.148593864053382	0.0047945879819782\\
0.16301159119982	0.00448295568132748\\
0.178451262965258	0.00418496841010679\\
0.194931728487138	0.00389980128576572\\
0.212480946909287	0.00362637289437857\\
0.231120822824294	0.00336386758601882\\
0.250978639035745	0.00310978897906043\\
0.271978464501058	0.00286561524637108\\
0.294131385231194	0.00263095577384581\\
0.317439921595682	0.0024055944818181\\
0.341914522738008	0.00218925128997719\\
0.367536121850307	0.00198205032375371\\
0.39429154969659	0.00178401651367482\\
0.422159059147457	0.0015952629238918\\
0.451863075985602	0.00141003303950409\\
0.482701456183384	0.00123504897358262\\
0.514633117213423	0.0010705392500333\\
0.547608060326365	0.000916771512477177\\
0.581578554129258	0.000773977602828258\\
0.616476161659445	0.000642479901824173\\
0.652234624389173	0.000522547715881464\\
0.688781512715622	0.000414445443339828\\
0.72607629657833	0.00031831094785264\\
0.764004615895544	0.000234506969756156\\
0.802482712777405	0.000163235765848844\\
0.841423246430145	0.000104674431688577\\
0.880738453533623	5.89679897096857e-05\\
0.920335111801815	2.62362869437555e-05\\
0.960120318204713	6.56332063028342e-06\\
1	0\\
};
\addlegendentry{Star graph, $d_{geo(\bar{\textbf{S}})}^2$};

\addplot [color=blue,only marks,mark=o,mark options={solid}]
  table[row sep=crcr]{%
0	0.398646427431691\\
0.000287555516361396	0.388521507068438\\
0.00115024114080624	0.378459363722443\\
0.00258809094009515	0.368459633489378\\
0.00460104235481953	0.358522760029815\\
0.00718916080201597	0.348648750274334\\
0.0103519441113652	0.338839759868234\\
0.0140888563324138	0.329097432926348\\
0.0183990602489732	0.319423933820657\\
0.0232839453888697	0.309817097047365\\
0.0287401871830403	0.300284285604373\\
0.0347661037377773	0.290828419104449\\
0.0413596136762872	0.281452746621222\\
0.0485186070580042	0.272160297989023\\
0.0562400635923893	0.262955023169496\\
0.0645208735912393	0.253840615118002\\
0.0733574907686649	0.244820964603989\\
0.082764908592223	0.235882647308715\\
0.0927220867189419	0.227047530737189\\
0.103224349889692	0.218319753463\\
0.114266579427194	0.209703560871679\\
0.125843519982743	0.201203058616829\\
0.137949089600535	0.192822719747529\\
0.150577051312132	0.184566855566432\\
0.163720740320821	0.176439817981894\\
0.177374705214316	0.168445041677798\\
0.191530089475426	0.160587786736233\\
0.206179141448758	0.15287235195619\\
0.221313707639024	0.14530302841894\\
0.236925449641501	0.137883988071214\\
0.25300537319469	0.130619508219607\\
0.269544300276261	0.123513732996109\\
0.286532683410823	0.11657076258395\\
0.30405121493773	0.109760095708092\\
0.322003543085638	0.10312200875906\\
0.340379010617048	0.0966604414574252\\
0.359166620132057	0.0903792603273809\\
0.378355183862324	0.0842822072367446\\
0.397933012671806	0.0783729985457855\\
0.417888238352491	0.0726552162366043\\
0.438208692821655	0.0671323468077369\\
0.458882623029126	0.0618075972963143\\
0.479896685597021	0.0566844199999266\\
0.501237946453033	0.051765968398447\\
0.522893214145056	0.04705528297602\\
0.544849137469913	0.0425552692983167\\
0.56709201749678	0.0382687365949676\\
0.589608012519914	0.0341983527461754\\
0.612383067787619	0.0303466589637989\\
0.635406271821938	0.0267155531122597\\
0.658659942987138	0.0233078852751277\\
0.682129491753567	0.0201257938816584\\
0.705800161246027	0.0171712801701918\\
0.729657086348385	0.0144462004406693\\
0.75368519680387	0.0119522770407826\\
0.777869334304864	0.00969108366005797\\
0.802194220176118	0.00766404920554135\\
0.826644658949516	0.00587244287326698\\
0.85120499905512	0.00431741468327178\\
0.875859691593012	0.00299994936959405\\
0.900593115103608	0.00192088192280992\\
0.925389603391731	0.0010808961520339\\
0.950233422417368	0.00048052534822306\\
0.97510881507099	0.000120150399337892\\
1	0\\
};
\addlegendentry{Star graph, $d_{diff}$};

\addplot [color=green,solid]
  table[row sep=crcr]{%
0	1\\
0.00187614314044537	0.918392699575457\\
0.00211865375990005	0.860126258409125\\
0.00252524207350093	0.809436414547463\\
0.00309137003480849	0.759741423826326\\
0.00382711188935687	0.711054754481474\\
0.00471959133262567	0.663641448279417\\
0.00576676763296879	0.617631809235044\\
0.00696637583030744	0.573153413703062\\
0.00833485993761662	0.530098632022053\\
0.00985154054103859	0.488817252584396\\
0.0115136728190808	0.449419279224968\\
0.0133186484982885	0.41200193222782\\
0.0152625728530759	0.376676220406929\\
0.0173450507040688	0.343475599337862\\
0.0195651426201116	0.312448272698345\\
0.0219242467398194	0.283601447685648\\
0.0243221261262452	0.258325989496944\\
0.0268579549674771	0.23481632872104\\
0.0295379735464662	0.213034903821763\\
0.0323697220566131	0.192934444776192\\
0.0353611176838049	0.174479472189306\\
0.0385229623865985	0.157600524989738\\
0.0418663121458655	0.142245835503553\\
0.0454044455688959	0.128353956995586\\
0.049137464404761	0.116008763761423\\
0.0530997032043817	0.10493213465966\\
0.0573107930677447	0.0950659574622452\\
0.0617933361953203	0.0863500610340032\\
0.0665680930690948	0.0787553140493249\\
0.0716721405066301	0.0721453650722345\\
0.0771435023329923	0.0664427095427208\\
0.0830328012623526	0.0615355548076063\\
0.0892509050377568	0.0572469421060032\\
0.096069358309031	0.053266501482086\\
0.103576921330938	0.0495279626360707\\
0.11182950768851	0.0460105860193001\\
0.120918235026198	0.0426908915707727\\
0.130876046584743	0.0395582739445054\\
0.141758265717875	0.0365982706703749\\
0.153623613993206	0.0337980534418864\\
0.166587007427289	0.0311397358970082\\
0.180688426307299	0.0286169682967247\\
0.196007254764767	0.0262216532716681\\
0.21262900737594	0.0239468620894027\\
0.230657065616816	0.0217861316905263\\
0.250184629495087	0.0197339240787435\\
0.271317702105264	0.0177856719743847\\
0.294160057205698	0.0159375651789616\\
0.321356627718732	0.0140973783041702\\
0.351015026920899	0.0123606496707269\\
0.383232103840568	0.010726264259682\\
0.418032893619751	0.00919570236793668\\
0.455496415585692	0.00776972305911122\\
0.495486131304697	0.00645342031930914\\
0.537905095093549	0.00525030147956193\\
0.582604850907775	0.00416408240865558\\
0.629785319426926	0.00319761913151716\\
0.678906416240891	0.00235463732387858\\
0.729733832230266	0.00163781718466022\\
0.782008238453601	0.00104921447125699\\
0.835465478362924	0.000590663247314022\\
0.889812484363043	0.000262629150197745\\
0.944756506971824	6.56730184061034e-05\\
1	0\\
};
\addlegendentry{Ring graph, $d_{geo(\bar{\textbf{S}})}^2$};

\addplot [color=green,only marks,mark=o,mark options={solid}]
  table[row sep=crcr]{%
0	1\\
0.00267968787534386	0.964450783920467\\
0.00352063518241092	0.938106564585437\\
0.00492956439058954	0.911028891933685\\
0.00688343656505721	0.88368000134514\\
0.00941286768078508	0.855977931693726\\
0.0124652792490692	0.828359639724664\\
0.0160266935143704	0.80097188922288\\
0.0200837229546033	0.773928870774548\\
0.0246552144338127	0.747370411035726\\
0.0296938087308533	0.721304429468603\\
0.0351884516622801	0.695784295222273\\
0.0411309136303553	0.670836549258232\\
0.0475102705657099	0.646497781319107\\
0.0543264615318356	0.622744766119672\\
0.0615770914394981	0.599573185811566\\
0.069262202498359	0.576971479069516\\
0.0772830868273642	0.555337472271519\\
0.0857416834029944	0.534189192594166\\
0.0946416076217671	0.513510504626182\\
0.103987201104863	0.493284851630411\\
0.113783302700262	0.473496787078035\\
0.124035693292288	0.454128831100748\\
0.134750313083639	0.435165431391008\\
0.145933091165211	0.416592427199869\\
0.157602625289307	0.398378801502576\\
0.169753415299531	0.380533011553112\\
0.182390349838626	0.363046370987322\\
0.195516728411714	0.345913419540758\\
0.209137830531937	0.329127150151293\\
0.223252034284133	0.312689928643776\\
0.237859273161555	0.29660211100386\\
0.252957465045527	0.280866473736834\\
0.269541516901873	0.264658328183789\\
0.286708745146112	0.248845349260081\\
0.304453039315723	0.233433824947189\\
0.322765805155018	0.218431806316643\\
0.341639503783035	0.203846401321233\\
0.361060488617834	0.189688560788388\\
0.38101610825644	0.175967903973377\\
0.401492077311976	0.162694511694006\\
0.422484113013793	0.149873518695395\\
0.443965777167979	0.137520232977027\\
0.465919893651356	0.125644542467916\\
0.48832823985325	0.114256191019715\\
0.511172480434354	0.103364491178305\\
0.534432497944852	0.0929786424862025\\
0.558088119788861	0.0831073319060065\\
0.582118578092571	0.0737588028295935\\
0.606515873118881	0.0649463111148075\\
0.631247309779335	0.05667074360314\\
0.656290886145769	0.0489388796832955\\
0.681624356929892	0.0417569267071719\\
0.707224991861732	0.0351307697803111\\
0.733070488518335	0.0290653009732166\\
0.75913813262606	0.0235651243550576\\
0.785405291257053	0.0186342445486281\\
0.811845403590204	0.0142776327599062\\
0.838439943146353	0.0104963675300491\\
0.865166385978547	0.00729295898385237\\
0.892002472098143	0.00466941450966774\\
0.918925564536285	0.00262740326252277\\
0.945914110084499	0.00116800177302986\\
0.972946178949074	0.000292040418438926\\
1	0\\
};
\addlegendentry{Ring graph, $d_{diff}$};

\end{axis}
\end{tikzpicture}%

%% file: gaussianKernel.tex
%
%
\begin{tikzpicture}

\begin{axis}[%
width=4.5in,
height=2.5in,
at={(0.911111in,0.513333in)},
scale only axis,
every outer x axis line/.append style={black},
every x tick label/.append style={font=\color{black}},
xmin=0,
xmax=1,
xlabel={$\Delta_s$},
every outer y axis line/.append style={black},
every y tick label/.append style={font=\color{black}},
ymin=0,
ymax=1,
ytick={0,0.2,0.4,0.6,0.8,1.0},
ylabel={$\Delta_{\mathcal{G}, u_0}$ (normalized)},
axis x line*=bottom,
axis y line*=left,
legend style={legend cell align=left,align=left,draw=black}
]
\addplot [color=red,solid]
  table[row sep=crcr]{%
0	1\\
0.000818313547723747	0.964242689346978\\
0.00145172635165655	0.931408256244913\\
0.00251057139528383	0.899073077748049\\
0.00399107417737345	0.867268939817533\\
0.00591266849284793	0.83593728762058\\
0.00825538140026858	0.805170426948255\\
0.0110186133985751	0.774983399987386\\
0.0142017686551792	0.745389657374408\\
0.017827174109269	0.716337668730592\\
0.0218724196101167	0.68790339065914\\
0.0263370483961929	0.660097312667211\\
0.0312208229187694	0.632928086032284\\
0.0365232551511123	0.606405950325942\\
0.0422449134914466	0.580534424514681\\
0.0483862643308397	0.555318894603989\\
0.0549482255007526	0.530762926188928\\
0.0619197550262803	0.506993323011737\\
0.0693132988548653	0.483872638878221\\
0.0771308177474564	0.461401711869702\\
0.085374625054037	0.439580067217382\\
0.0940471050415828	0.4184086748307\\
0.103151527039875	0.39788284147098\\
0.112691292822608	0.377999533959136\\
0.122670399688504	0.358754074555682\\
0.133088494844305	0.340162357141678\\
0.143955497738961	0.322193232206872\\
0.155277151353057	0.304839695291525\\
0.1670601818938	0.288093060959576\\
0.17931064954172	0.271947692952395\\
0.192039021469358	0.256387462399987\\
0.205255670543442	0.241399866207426\\
0.21897381482721	0.226969095487639\\
0.233553494893149	0.21321239666132\\
0.248714990073292	0.199933673267707\\
0.264476397924089	0.187118792950405\\
0.280854014616916	0.174755827333517\\
0.297869147896737	0.162832123286133\\
0.315530531704495	0.151341731546688\\
0.333850154667473	0.140277899905883\\
0.352835836614805	0.129636291283718\\
0.372542068245117	0.119399497001348\\
0.392929876922982	0.109580474254244\\
0.414000337080205	0.100178209494217\\
0.435751177965732	0.0911927134817097\\
0.458181789924534	0.0826240099657449\\
0.481283744854028	0.0744730915895877\\
0.505050359479654	0.0667408430673121\\
0.529474561933434	0.0594276964545516\\
0.555042729057005	0.0524836382385176\\
0.581296063279626	0.0459535863159041\\
0.608224192674181	0.0398386082710507\\
0.635808249051725	0.0341406671917823\\
0.664042145981206	0.0288605759835898\\
0.692874543206009	0.0240040429726057\\
0.72226274619361	0.0195756474870683\\
0.752146197250847	0.0155818655279602\\
0.782698349487149	0.0120125748366671\\
0.813608426790815	0.00889116260484967\\
0.844775686178167	0.00622547774163776\\
0.876100005381786	0.00402219899195317\\
0.907674295149656	0.00228351325959882\\
0.939150550269622	0.00102645717871205\\
0.970316579854103	0.000260019319048694\\
1	0\\
};
\addlegendentry{Random geometric graph, $d_{geo(\textbf{E})}^2$};

\addplot [color=red,only marks,mark=o,mark options={solid}]
  table[row sep=crcr]{%
0	1\\
0.000335371833811698	0.980433515034006\\
0.00134174326614596	0.96070597306764\\
0.00301978423554376	0.94082115358181\\
0.0053701823210551	0.920783226255129\\
0.00839448627426184	0.900597395754703\\
0.0120926187881648	0.880267830593815\\
0.0164648194496057	0.859800857079426\\
0.0215106226552146	0.839203995165306\\
0.0272409668822603	0.818476773086248\\
0.0336449960437102	0.797635789454695\\
0.0407200610143738	0.776693109633111\\
0.0484620561275944	0.755662483105456\\
0.0568674367061377	0.734558188878089\\
0.0659289287852137	0.713395094879214\\
0.0756395848458025	0.692188611021882\\
0.0859912420611529	0.670954446077245\\
0.0970743681455927	0.64961880776945\\
0.108794366743491	0.62827996084753\\
0.12114003535119	0.606956121252976\\
0.134098555079658	0.585667318399919\\
0.147657420960279	0.56443273901253\\
0.161800196514509	0.543276405633857\\
0.176510600473909	0.522221279966419\\
0.191770955640265	0.501291393394635\\
0.207569702429056	0.480504723618155\\
0.223881250731204	0.459891799842375\\
0.240685157895838	0.439477246881221\\
0.257960087594419	0.419285566226556\\
0.275684236339178	0.399341228304226\\
0.293834910275413	0.379667471370131\\
0.312389003390161	0.360287440251875\\
0.331322975393312	0.341223519488177\\
0.350952311326053	0.322285273185615\\
0.370947007719743	0.303695057974356\\
0.391280278289864	0.285475640268693\\
0.411924705894149	0.267649508301331\\
0.432852452486482	0.250238721828128\\
0.454035118139902	0.233264958423625\\
0.475443972053715	0.216749376435347\\
0.497050000369325	0.200712559109708\\
0.518822737789647	0.185175427450131\\
0.54073411223634	0.170156251559572\\
0.562754591351502	0.15567369069438\\
0.584854666566371	0.141745648449819\\
0.607004512529279	0.128389471016688\\
0.629175037731679	0.115621383944771\\
0.651336932094712	0.103457026628628\\
0.673461182946698	0.091911187844017\\
0.695485721896367	0.0810113909354239\\
0.717416282005782	0.0707555003643347\\
0.739224867197411	0.0611557539372597\\
0.760884056088563	0.0522234721412608\\
0.782366228839907	0.0439693148345602\\
0.803645694867002	0.036402598336338\\
0.824696658535249	0.0295319932768292\\
0.845494240136442	0.0233652044809346\\
0.866005979703953	0.0179108696182069\\
0.886217244557674	0.0131725686854747\\
0.906105626199198	0.00915533245343776\\
0.925650046074677	0.00586325895730555\\
0.944828917186207	0.00329977831596808\\
0.963625006024959	0.00146708038929657\\
0.982020629624346	0.000366845101118913\\
1	0\\
};
\addlegendentry{Random geometric graph, $d_{diff}$};

\end{axis}
\end{tikzpicture}%

%% file: cameraman.tex
%
%
\begin{tikzpicture}

\begin{axis}[%
width=3.5in,
height=2.in,
at={(0.911111in,0.513333in)},
scale only axis,
every outer x axis line/.append style={black},
every x tick label/.append style={font=\color{black}},
xmin=0,
xmax=1,
xlabel={$\Delta_s$},
every outer y axis line/.append style={black},
every y tick label/.append style={font=\color{black}},
ymin=0,
ymax=1,
xtick={0,0.2,0.4,0.6,0.8,1.0},
ytick={0,0.2,0.4,0.6,0.8,1.0},
ylabel={$\Delta_{\mathcal{G}, u_0}$ (normalized)},
axis x line*=bottom,
axis y line*=left,
legend style={legend cell align=left,align=left,draw=black}
]
\addplot [color=red,solid]
  table[row sep=crcr]{%
0	1\\
3.2328747074077e-05	0.920971695213793\\
8.514228351064e-05	0.862926863656398\\
0.000170953011937279	0.799598915915133\\
0.000279152732423621	0.740122573904403\\
0.000391023225894473	0.690486232529846\\
0.000564895304658284	0.627054100778505\\
0.000724910169118186	0.57820570007784\\
0.00104265097790432	0.498619043134403\\
0.00107390312759826	0.491743294242302\\
0.00134260963279337	0.438062135973422\\
0.00148793878851118	0.412521599776294\\
0.00202323107520969	0.334688811405381\\
0.00229302507083861	0.305304560143442\\
0.00256281906646753	0.275920308881502\\
0.00284346555560093	0.251289085206353\\
0.00396365216240075	0.180964804979591\\
0.0041415821738169	0.173086441008203\\
0.00431951218523305	0.165208077036815\\
0.00451952845279326	0.157332419124569\\
0.00574799568157694	0.120265788401436\\
0.00588524115453734	0.117050825529305\\
0.00699900985111452	0.0956613220165033\\
0.00779149375829102	0.0843160744630496\\
0.00953389588190912	0.0665743472443284\\
0.0103363749692011	0.0606601684650784\\
0.0117590450668241	0.0524344217568237\\
0.0131618980427519	0.0462963597592532\\
0.0150624069958028	0.0400112574664633\\
0.0169639772829645	0.0352570699505006\\
0.0192458849847538	0.0308697104110576\\
0.0218219584834011	0.0270650686089608\\
0.0247803588087464	0.0236969701635305\\
0.0277821576327843	0.0210248330719913\\
0.0311400901865294	0.0186531424322497\\
0.0349179037102746	0.0165350575213481\\
0.0391340895041089	0.0146579217965376\\
0.0438915922783023	0.0129762310749729\\
0.0492050056210091	0.0114855682570027\\
0.0551777472467639	0.0101556136598882\\
0.0618540922517749	0.00897579409436387\\
0.0693801603629962	0.00792075810736438\\
0.0778034530606692	0.00698459360257738\\
0.087267198555584	0.00615088428385289\\
0.0978677910498379	0.00541091561075534\\
0.109790616832595	0.00475179042750236\\
0.123156280964945	0.00416701753632456\\
0.138167494712808	0.00364757333845869\\
0.155000629701831	0.00318729479290547\\
0.17395966574329	0.00277828632931409\\
0.195230468452199	0.00241695331232198\\
0.219123183426127	0.00209794192157122\\
0.24602084063493	0.00181609808115538\\
0.276344543138138	0.00156653927134211\\
0.310955755643664	0.00134175264119706\\
0.3508309041172	0.00113545592439529\\
0.396679064501508	0.000943981796202929\\
0.455527306709665	0.000745237463764933\\
0.522068194830844	0.000563693873729895\\
0.595455239896024	0.000401674476040822\\
0.674016592901762	0.0002629057098572\\
0.75620795613244	0.000150305561122871\\
0.839460905710964	6.75868436116556e-05\\
0.921490315300701	1.70123491841065e-05\\
1	0\\
};
\addlegendentry{Cameraman image graph, $d_{geo(\bar{\textbf{S}})}^2$};

\addplot [color=red,only marks,mark=o,mark options={solid}]
  table[row sep=crcr]{%
0	1\\
0.000306152974065852	0.989775192070554\\
0.0018630616658143	0.967426156258857\\
0.00395888884548802	0.944540259668909\\
0.00819167466740217	0.906678597962038\\
0.0109054968005724	0.886088764295503\\
0.0151681621221403	0.857960163009081\\
0.0193422450013348	0.834218713160556\\
0.0246526035437656	0.807961092062642\\
0.0293881152708281	0.787300841033361\\
0.0348162214924826	0.766010418319803\\
0.0406237208221282	0.745417650135736\\
0.0469918390797683	0.724864583762489\\
0.0537888248345977	0.704780815337287\\
0.0610936062132628	0.68490671944504\\
0.0689173196337919	0.665213899364717\\
0.0772420213192308	0.645741766728075\\
0.0863399124260909	0.625902703490517\\
0.0959718618323169	0.606263153960445\\
0.106195063477593	0.586703017120201\\
0.116958860480322	0.567316066741963\\
0.128354376291974	0.547939986427343\\
0.14029436767094	0.528725234341806\\
0.152834162554693	0.509580168634762\\
0.165911152695692	0.490598270150202\\
0.179650117137163	0.47160361339452\\
0.19391683102184	0.452787051049865\\
0.208759692089555	0.43408288133485\\
0.224107242546344	0.415579590473686\\
0.240037307088071	0.397184180333696\\
0.256444395043696	0.379019456359217\\
0.27337080087046	0.361037172647483\\
0.290738638228496	0.343318852978514\\
0.309906144426587	0.324557446249068\\
0.329539651529167	0.306136730708765\\
0.349671095980219	0.28802553103562\\
0.370213854007184	0.270300733968873\\
0.391226754768799	0.252912065756831\\
0.412593511097305	0.235956360101104\\
0.434339652875314	0.219412370658852\\
0.456378125149087	0.20334547989659\\
0.478773271466259	0.187709815775466\\
0.501398404549112	0.172594971609087\\
0.524272308034002	0.157987582216419\\
0.547311710502251	0.143940138521202\\
0.570555568900066	0.130429152628765\\
0.593900632473383	0.117515026787004\\
0.61735841327362	0.105190719272194\\
0.640853359335498	0.0934953789678908\\
0.664386671709172	0.0824275318877954\\
0.687894466078682	0.0720164772316526\\
0.711380649975682	0.0622599416466536\\
0.734780613970031	0.053184353433407\\
0.758108352091383	0.0447843868148539\\
0.781291470425994	0.0370863473875376\\
0.804326216685552	0.0300910639154688\\
0.82716096975495	0.0238139322037325\\
0.849793198456432	0.0182555059971706\\
0.872173377635285	0.0134279479956486\\
0.894290873485113	0.00933326257769488\\
0.916108206255101	0.00597824195641772\\
0.937614990505419	0.00336441351071036\\
0.958777707489928	0.00149592994525451\\
0.979580406209422	0.000374045138332795\\
1	0\\
};
\addlegendentry{Cameraman image graph, $d_{diff}$};

\end{axis}
\end{tikzpicture}%